\documentclass[oneside,reqno,american]{amsart}
\usepackage[T1]{fontenc}
\usepackage{verbatim}
\usepackage{amsbsy}
\usepackage{amstext}
\usepackage{amsthm}
\usepackage{amssymb}
\usepackage{undertilde}
\usepackage{graphicx}
\usepackage{xargs}[2008/03/08]

\makeatletter
\numberwithin{equation}{section}
\numberwithin{figure}{section}
\theoremstyle{plain}
\newtheorem{thm}{\protect\theoremname}[section]
\theoremstyle{remark}
\newtheorem{rem}[thm]{\protect\remarkname}

\usepackage{pdfsync}

\usepackage[scale=1.05,lining,tabular]{fbb} 
\usepackage[stix2,bigdelims,vvarbb,slantedGreek]{newtxmath}
\usepackage[cal=boondoxo,bb=boondox,frak=boondox]{mathalfa}

\makeatother

\usepackage{babel}
\providecommand{\remarkname}{Remark}
\providecommand{\theoremname}{Theorem}

\begin{document}

\global\long\def\ga{\alpha}%
\global\long\def\gb{\beta}%
\global\long\def\ggm{\gamma}%
\global\long\def\go{\omega}%
\global\long\def\ge{\epsilon}%
\global\long\def\gs{\sigma}%
\global\long\def\gd{\delta}%
\global\long\def\gD{\Delta}%
\global\long\def\vph{\varphi}%
\global\long\def\gf{\varphi}%
\global\long\def\gk{\kappa}%
\global\long\def\gl{\lambda}%
\global\long\def\nb{\boldsymbol{\nabla}}%
\global\long\def\scp{\boldsymbol{\cdot}}%

\global\long\def\eps{\varepsilon}%
\global\long\def\epss#1#2{\varepsilon_{#2}^{#1}}%
\global\long\def\ep#1{\eps_{#1}}%

\global\long\def\wh#1{\widehat{#1}}%

\global\long\def\spec#1{\textsf{#1}}%

\global\long\def\ui{\wh{\boldsymbol{\imath}}}%
\global\long\def\uj{\wh{\boldsymbol{\jmath}}}%
\global\long\def\uk{\widehat{\boldsymbol{k}}}%

\global\long\def\uI{\widehat{\mathbf{I}}}%
\global\long\def\uJ{\widehat{\mathbf{J}}}%
\global\long\def\uK{\widehat{\mathbf{K}}}%

\global\long\def\bs#1{\boldsymbol{#1}}%
\global\long\def\vect#1{\mathbf{#1}}%
\global\long\def\bi#1{\textbf{\emph{#1}}}%

\global\long\def\uv#1{\widehat{\boldsymbol{#1}}}%
\global\long\def\cross{\bs{\times}}%

\global\long\def\ddt{\frac{\dee}{\dee t}}%
\global\long\def\dbyd#1{\frac{\dee}{\dee#1}}%
\global\long\def\dby#1#2{\frac{\partial#1}{\partial#2}}%

\global\long\def\vct#1{\boldsymbol{#1}}%

\global\long\def\partialby#1#2{\frac{\partial#1}{\partial x^{#2}}}%
\newcommandx\parder[2][usedefault, addprefix=\global, 1=]{\frac{\partial#2}{\partial#1}}%

\global\long\def\oneto{1,\dots,}%
\global\long\def\mi#1{\boldsymbol{#1}}%
\global\long\def\mii{\mi I}%

\global\long\def\fall{,\quad\text{for all}\quad}%

\global\long\def\reals{\mathbb{R}}%

\global\long\def\rthree{\reals^{3}}%
\global\long\def\rsix{\reals^{6}}%
\global\long\def\rn{\reals^{n}}%
\global\long\def\rt#1{\reals^{#1}}%

\global\long\def\les{\leqslant}%
\global\long\def\ges{\geqslant}%

\global\long\def\dee{\textrm{d}}%
\global\long\def\di{d}%

\global\long\def\from{\colon}%
\global\long\def\tto{\longrightarrow}%
\global\long\def\lmt{\longmapsto}%

\global\long\def\abs#1{\left|#1\right|}%

\global\long\def\isom{\cong}%

\global\long\def\comp{\circ}%

\global\long\def\cl#1{\overline{#1}}%

\global\long\def\fun{\varphi}%

\global\long\def\interior{\textrm{Int}\,}%

\global\long\def\sign{\textrm{sign}\,}%
\global\long\def\sgn#1{(-1)^{#1}}%
\global\long\def\sgnp#1{(-1)^{\abs{#1}}}%

\global\long\def\dimension{\textrm{dim}\,}%

\global\long\def\esssup{\textrm{ess}\,\sup}%

\global\long\def\ess{\textrm{{ess}}}%

\global\long\def\kernel{\mathop{\textrm{Kernel}}}%

\global\long\def\support{\textrm{supp}\,}%

\global\long\def\image{\textrm{Image}\,}%

\global\long\def\diver{\mathop{\textrm{div}}}%

\global\long\def\sp{\mathop{\textrm{span}}}%

\global\long\def\resto#1{|_{#1}}%
\global\long\def\incl{\iota}%
\global\long\def\iden{\imath}%
\global\long\def\idnt{\textrm{Id}}%
\global\long\def\rest{\rho}%
\global\long\def\extnd{e_{0}}%

\global\long\def\proj{\textrm{pr}}%

\global\long\def\ino#1{\int_{#1}}%

\global\long\def\half{\frac{1}{2}}%
\global\long\def\shalf{{\scriptstyle \half}}%
\global\long\def\third{\frac{1}{3}}%

\global\long\def\empt{\varnothing}%

\global\long\def\paren#1{\left(#1\right)}%
\global\long\def\bigp#1{\bigl(#1\bigr)}%
\global\long\def\biggp#1{\biggl(#1\biggr)}%
\global\long\def\Bigp#1{\Bigl(#1\Bigr)}%

\global\long\def\braces#1{\left\{  #1\right\}  }%
\global\long\def\sqbr#1{\left[#1\right]}%
\global\long\def\anglep#1{\left\langle #1\right\rangle }%

\global\long\def\lsum{{\textstyle \sum}}%

\global\long\def\bigabs#1{\bigl|#1\bigr|}%

\global\long\def\lisub#1#2#3{#1_{1}#2\dots#2#1_{#3}}%

\global\long\def\lisup#1#2#3{#1^{1}#2\dots#2#1^{#3}}%

\global\long\def\lisubb#1#2#3#4{#1_{#2}#3\dots#3#1_{#4}}%

\global\long\def\lisubbc#1#2#3#4{#1_{#2}#3\cdots#3#1_{#4}}%

\global\long\def\lisubbwout#1#2#3#4#5{#1_{#2}#3\dots#3\widehat{#1}_{#5}#3\dots#3#1_{#4}}%

\global\long\def\lisubc#1#2#3{#1_{1}#2\cdots#2#1_{#3}}%

\global\long\def\lisupc#1#2#3{#1^{1}#2\cdots#2#1^{#3}}%

\global\long\def\lisupp#1#2#3#4{#1^{#2}#3\dots#3#1^{#4}}%

\global\long\def\lisuppc#1#2#3#4{#1^{#2}#3\cdots#3#1^{#4}}%

\global\long\def\lisuppwout#1#2#3#4#5#6{#1^{#2}#3#4#3\wh{#1^{#6}}#3#4#3#1^{#5}}%

\global\long\def\lisubbwout#1#2#3#4#5#6{#1_{#2}#3#4#3\wh{#1}_{#6}#3#4#3#1_{#5}}%

\global\long\def\lisubwout#1#2#3#4{#1_{1}#2\dots#2\widehat{#1}_{#4}#2\dots#2#1_{#3}}%

\global\long\def\lisupwout#1#2#3#4{#1^{1}#2\dots#2\widehat{#1^{#4}}#2\dots#2#1^{#3}}%

\global\long\def\lisubwoutc#1#2#3#4{#1_{1}#2\cdots#2\widehat{#1}_{#4}#2\cdots#2#1_{#3}}%

\global\long\def\twp#1#2#3{\dee#1^{#2}\wedge\dee#1^{#3}}%

\global\long\def\thp#1#2#3#4{\dee#1^{#2}\wedge\dee#1^{#3}\wedge\dee#1^{#4}}%

\global\long\def\fop#1#2#3#4#5{\dee#1^{#2}\wedge\dee#1^{#3}\wedge\dee#1^{#4}\wedge\dee#1^{#5}}%

\global\long\def\idots#1{#1\dots#1}%
\global\long\def\icdots#1{#1\cdots#1}%

\global\long\def\norm#1{\|#1\|}%

\global\long\def\nonh{\heartsuit}%

\global\long\def\nhn#1{\norm{#1}^{\nonh}}%

\global\long\def\trps{^{{\scriptscriptstyle \textsf{T}}}}%

\global\long\def\testfuns{\mathcal{D}}%

\global\long\def\ntil#1{\tilde{#1}{}}%

\global\long\def\alt{\mathfrak{A}}%

\global\long\def\pou{\eta}%

\global\long\def\ext{{\textstyle \bigwedge}}%
\global\long\def\forms{\Omega}%

\global\long\def\dotwedge{\dot{\mbox{\ensuremath{\wedge}}}}%

\global\long\def\vel{\theta}%

\global\long\def\contr{\raisebox{0.4pt}{\mbox{\ensuremath{\lrcorner}}}\,}%
\global\long\def\fcontr{\raisebox{0.4pt}{\mbox{\ensuremath{\llcorner}}}\,}%

\global\long\def\lie{\mathcal{L}}%

\global\long\def\L#1{L\bigl(#1\bigr)}%

\global\long\def\vvforms{\ext^{\dims}\bigp{T\spc,\vbts^{*}}}%

\global\long\def\spc{\mathcal{S}}%
\global\long\def\sptm{\mathcal{E}}%
\global\long\def\evnt{e}%
\global\long\def\frame{\Phi}%

\global\long\def\timeman{\mathcal{T}}%
\global\long\def\zman{t}%
\global\long\def\dims{n}%
\global\long\def\m{\dims-1}%
\global\long\def\dimw{m}%

\global\long\def\wc{z}%

\global\long\def\fourv#1{\mbox{\ensuremath{\mathfrak{#1}}}}%

\global\long\def\pbform#1{\utilde{#1}}%
\global\long\def\util#1{\raisebox{-5pt}{\ensuremath{{\scriptscriptstyle \sim}}}\!\!\!#1}%

\global\long\def\utilJ{\util J}%

\global\long\def\utilRho{\util{\rho}}%

\global\long\def\body{B}%
\global\long\def\man{\mathcal{M}}%
\global\long\def\var{\mathcal{V}}%

\global\long\def\bdry{\partial}%

\global\long\def\gO{\varOmega}%

\global\long\def\reg{\mathcal{R}}%
\global\long\def\bdrr{\bdry\reg}%

\global\long\def\bdom{\bdry\gO}%

\global\long\def\bndo{\partial\gO}%

\global\long\def\base{\mathcal{X}}%
\global\long\def\fb{\mathcal{Y}}%

\global\long\def\pform{\varsigma}%
\global\long\def\vform{\beta}%
\global\long\def\sform{\tau}%
\global\long\def\flow{J}%
\global\long\def\fv{\vct h}%
\global\long\def\n{\m}%
\global\long\def\cmap{\mathfrak{t}}%
\global\long\def\vcmap{\varSigma}%

\global\long\def\mvec{\mathfrak{v}}%
\global\long\def\mveco#1{\mathfrak{#1}}%
\global\long\def\smbase{\mathfrak{e}}%
\global\long\def\spx{\simp}%

\global\long\def\hp{H}%
\global\long\def\ohp{h}%

\global\long\def\hps{G_{\dims-1}(T\spc)}%
\global\long\def\ohps{G_{\dims-1}^{\perp}(T\spc)}%
\global\long\def\hpsx{G_{\dims-1}(\tspc)}%
\global\long\def\ohpsx{G_{\dims-1}^{\perp}(\tspc)}%

\global\long\def\fbun{F}%

\global\long\def\flowm{\Phi}%

\global\long\def\tgb{T\spc}%
\global\long\def\ctgb{T^{*}\spc}%
\global\long\def\tspc{T_{\pis}\spc}%
\global\long\def\dspc{T_{\pis}^{*}\spc}%

\global\long\def\fflow{\fourv J}%
\global\long\def\fvform{\mathfrak{b}}%
\global\long\def\fsform{\mathfrak{t}}%
\global\long\def\fpform{\mathfrak{s}}%

\global\long\def\maxw{\mathfrak{g}}%
\global\long\def\frdy{\mathfrak{f}}%
\global\long\def\ptnl{A}%

\global\long\def\eucl{E}%

\global\long\def\mind{\alpha}%
\global\long\def\vb{\xi}%

\global\long\def\man{\mathcal{M}}%
\global\long\def\odman{\mathcal{N}}%
\global\long\def\subman{\mathcal{A}}%

\global\long\def\vbt{\mathcal{E}}%
\global\long\def\fib{\mathbf{V}}%
\global\long\def\vbts{W}%
\global\long\def\avb{U}%

\global\long\def\chart{\varphi}%
\global\long\def\vbchart{\Phi}%

\global\long\def\jetb#1{J^{#1}}%
\global\long\def\jet#1{j^{1}(#1)}%
\global\long\def\tjet{\tilde{\jmath}}%

\global\long\def\Jet#1{J^{1}(#1)}%

\global\long\def\jetm#1{j_{#1}}%

\global\long\def\sobp#1#2{W_{#2}^{#1}}%

\global\long\def\inner#1#2{\left\langle #1,#2\right\rangle }%

\global\long\def\fields{\sobp pk(\vb)}%

\global\long\def\bodyfields{\sobp p{k_{\partial}}(\vb)}%

\global\long\def\forces{\sobp pk(\vb)^{*}}%

\global\long\def\bfields{\sobp p{k_{\partial}}(\vb\resto{\bndo})}%

\global\long\def\loadp{(\sfc,\bfc)}%

\global\long\def\strains{\lp p(\jetb k(\vb))}%

\global\long\def\stresses{\lp{p'}(\jetb k(\vb)^{*})}%

\global\long\def\diffop{D}%

\global\long\def\strainm{E}%

\global\long\def\incomps{\vbts_{\yieldf}}%

\global\long\def\devs{L^{p'}(\eta_{1}^{*})}%

\global\long\def\incompsns{L^{p}(\eta_{1})}%

\global\long\def\testf{\mathcal{D}}%
\global\long\def\dists{\mathcal{D}'}%

\global\long\def\codiv{\boldsymbol{\partial}}%

\global\long\def\currof#1{\tilde{#1}}%

\global\long\def\chn{c}%
\global\long\def\chnsp{\mathbf{F}}%

\global\long\def\current{T}%
\global\long\def\curr{R}%

\global\long\def\gdiv{\bdry\textrm{iv\,}}%

\global\long\def\prop{P}%

\global\long\def\aprop{Q}%

\global\long\def\flux{\omega}%
\global\long\def\aflux{S}%

\global\long\def\fform{\tau}%

\global\long\def\dimn{n}%

\global\long\def\sdim{{\dimn-1}}%

\global\long\def\contrf{{\scriptstyle \smallfrown}}%

\global\long\def\prodf{{\scriptstyle \smallsmile}}%

\global\long\def\ptnl{\varphi}%

\global\long\def\form{\omega}%

\global\long\def\dens{\rho}%

\global\long\def\simp{s}%
\global\long\def\ssimp{\Delta}%
\global\long\def\cpx{K}%

\global\long\def\cell{C}%

\global\long\def\chain{B}%

\global\long\def\ach{A}%

\global\long\def\coch{X}%

\global\long\def\scale{s}%

\global\long\def\fnorm#1{\norm{#1}^{\flat}}%

\global\long\def\chains{\mathcal{A}}%

\global\long\def\ivs{\boldsymbol{U}}%

\global\long\def\mvs{\boldsymbol{V}}%

\global\long\def\cvs{\boldsymbol{W}}%

\global\long\def\pis{x}%
\global\long\def\xo{\pis_{0}}%

\global\long\def\pib{X}%

\global\long\def\pbndo{\Gamma}%
\global\long\def\bndoo{\pbndo_{0}}%
 
\global\long\def\bndot{\pbndo_{t}}%

\global\long\def\cloo{\cl{\gO}}%

\global\long\def\nor{\vct n}%

\global\long\def\dA{\,\dee A}%

\global\long\def\dV{\,\dee V}%

\global\long\def\eps{\varepsilon}%

\global\long\def\vs{\mathbf{W}}%
\global\long\def\avs{\mathbf{V}}%
\global\long\def\affsp{\mathbf{A}}%
\global\long\def\pt{p}%

\global\long\def\vbase{e}%
\global\long\def\sbase{\mathbf{e}}%
\global\long\def\msbase{\mathfrak{e}}%
\global\long\def\vect{v}%

\global\long\def\vf{w}%

\global\long\def\avf{u}%

\global\long\def\stn{\varepsilon}%

\global\long\def\rig{r}%

\global\long\def\rigs{\mathcal{R}}%

\global\long\def\qrigs{\!/\!\rigs}%

\global\long\def\qd{\!/\,\!\kernel\diffop}%

\global\long\def\dis{\chi}%
\global\long\def\conf{\kappa}%
\global\long\def\confsp{\mathcal{Q}}%

\global\long\def\embds{\textrm{Emb}}%

\global\long\def\fc{F}%

\global\long\def\st{\sigma}%

\global\long\def\bfc{\vct b}%

\global\long\def\sfc{\vct t}%

\global\long\def\stm{S}%

\global\long\def\nhs{Y}%

\global\long\def\soc{Z}%

\global\long\def\tran{\mathrm{tr}}%

\global\long\def\slf{R}%

\global\long\def\sts{\varSigma}%

\global\long\def\ebdfc{T}%
\global\long\def\optimum{\st^{\textrm{opt}}}%
\global\long\def\scf{K}%

\global\long\def\cee#1{C^{#1}}%

\global\long\def\lone{L^{1}}%

\global\long\def\linf{L^{\infty}}%

\global\long\def\lp#1{L^{#1}}%

\global\long\def\ofbdo{(\bndo)}%

\global\long\def\ofclo{(\cloo)}%

\global\long\def\vono{(\gO,\rthree)}%

\global\long\def\vonbdo{(\bndo,\rthree)}%
\global\long\def\vonbdoo{(\bndoo,\rthree)}%
\global\long\def\vonbdot{(\bndot,\rthree)}%

\global\long\def\vonclo{(\cl{\gO},\rthree)}%

\global\long\def\strono{(\gO,\reals^{6})}%

\global\long\def\sob{W_{1}^{1}}%

\global\long\def\sobb{\sob(\gO,\rthree)}%

\global\long\def\lob{\lone(\gO,\rthree)}%

\global\long\def\lib{\linf(\gO,\reals^{12})}%

\global\long\def\ofO{(\gO)}%

\global\long\def\oneo{{1,\gO}}%
\global\long\def\onebdo{{1,\bndo}}%
\global\long\def\info{{\infty,\gO}}%

\global\long\def\infclo{{\infty,\cloo}}%

\global\long\def\infbdo{{\infty,\bndo}}%

\global\long\def\ld{LD}%

\global\long\def\ldo{\ld\ofO}%
\global\long\def\ldoo{\ldo_{0}}%

\global\long\def\trace{\gamma}%

\global\long\def\pr{\proj_{\rigs}}%

\global\long\def\pq{\proj}%

\global\long\def\qr{\,/\,\reals}%

\global\long\def\aro{S_{1}}%
\global\long\def\art{S_{2}}%

\global\long\def\mo{m_{1}}%
\global\long\def\mt{m_{2}}%

\global\long\def\yieldc{B}%

\global\long\def\yieldf{Y}%

\global\long\def\trpr{\pi_{P}}%

\global\long\def\devpr{\pi_{\devsp}}%

\global\long\def\prsp{P}%

\global\long\def\devsp{D}%

\global\long\def\ynorm#1{\|#1\|_{\yieldf}}%

\global\long\def\colls{\Psi}%

\global\long\def\ssx{S}%

\global\long\def\smap{s}%

\global\long\def\smat{\chi}%

\global\long\def\sx{e}%

\global\long\def\snode{P}%

\global\long\def\elem{e}%

\global\long\def\nel{L}%

\global\long\def\el{l}%

\global\long\def\ipln{\phi}%

\global\long\def\ndof{D}%

\global\long\def\dof{d}%

\global\long\def\nldof{N}%

\global\long\def\ldof{n}%

\global\long\def\lvf{\chi}%

\global\long\def\lfc{\varphi}%

\global\long\def\amat{A}%

\global\long\def\snomat{E}%

\global\long\def\femat{E}%

\global\long\def\tmat{T}%

\global\long\def\fvec{f}%

\global\long\def\snsp{\mathcal{S}}%

\global\long\def\slnsp{\Phi}%

\global\long\def\ro{r_{1}}%

\global\long\def\rtwo{r_{2}}%

\global\long\def\rth{r_{3}}%

\global\long\def\subbs{\mathcal{B}}%

\global\long\def\elements{\mathcal{E}}%

\global\long\def\element{E}%

\global\long\def\nodes{\mathcal{N}}%

\global\long\def\node{N}%

\global\long\def\psubbs{\mathcal{P}}%

\global\long\def\psubb{P}%

\global\long\def\matr{M}%

\global\long\def\nodemap{\nu}%

\global\long\def\B{\boldsymbol{B}}%
\global\long\def\H{\bs H}%
\global\long\def\J{\bs J}%
\global\long\def\E{\bs E}%
\global\long\def\D{\bs D}%
\global\long\def\A{\boldsymbol{A}}%
\global\long\def\bw{\vct w}%

\global\long\def\potl{\varphi}%

\global\long\def\ptnl{\alpha}%

\global\long\def\currsp{\mathcal{I}}%

\global\long\def\volt{V}%

\global\long\def\intv{\mathbf{t}}%
\global\long\def\intc{t}%
\global\long\def\intsp{\mathcal{T}}%

\global\long\def\frcv{\mathbf{f}}%
\global\long\def\frcc{f}%
\global\long\def\frcsp{\mathcal{F}}%

\global\long\def\velv{\mathbf{V}}%
\global\long\def\velc{V}%
\global\long\def\disv{\mathbf{E}}%
\global\long\def\disc{E}%

\global\long\def\posn{\mathbf{x}}%
\global\long\def\area{\mathbf{A}}%
\global\long\def\relp{\mathbf{L}}%

\global\long\def\chn{c}%

\title{Electrodynamics and Geometric Continuum Mechanics}
\author{Reuven Segev}
\address{Department of Mechanical Engineering, Ben-Gurion University of the
Negev, Beer-Sheva, Israel. rsegev@post.bgu.ac.il}
\date{\today}
\begin{abstract}
This paper offers an informal instructive introduction to some of
the main notions of geometric continuum mechanics for the case of
smooth fields. We use a metric invariant stress theory of continuum
mechanics to formulate a simple generalization of the fields of electrodynamics
and Maxwell's equations to general differentiable manifolds of any
dimension, thus viewing generalized electrodynamics as a special case
of continuum mechanics. The basic kinematic variable is the potential,
which is represented as a $p$-form in an $n$-dimensional spacetime.
The stress for the case of generalized electrodynamics is assumed
to be represented by an $(n-p-1)$-form, a generalization of the Maxwell
$2$-form.
\end{abstract}

\maketitle

\section{Introduction}

This paper offers an informal instructive introduction to some of
the main notions of geometric continuum mechanics for the case of
smooth fields. Usually, continuum mechanics is thought of as the theoretical
foundation for engineering stress analysis and fluid dynamics. Here,
in order to motivate the constructions of geometric continuum mechanics,
we show that the Maxwell equations of electromagnetism and generalizations
thereof to metric-independent, or premetric, $p$-form electrodynamics
in an $n$-dimensional spacetime may be obtained as a special case
of geometric continuum mechanics. (See \cite{Hehl2003,Kaiser2004,Hehl2006}
for premetric electromagnetism and \cite{Henneaux1986,Henneaux1988,Navarro2012}
for $p$-form electrodynamics.) For the non-smooth counterpart of
the theory, see \cite{Segev2016,Segev2023}.

The notations $\D,\,\E,\,\H,\,\B,\,\J$, and $\rho$, are used for
the electric displacement, electric field, magnetic field intensity,
magnetic field flux density, current density, and charge density,
respectively. Thus, the Maxwell equations are
\begin{alignat}{2}
\nb\scp\D & =\rho, &  & \qquad\text{Gauss's law,}\\
\nb\scp\B & =0, &  & \qquad\text{Gauss's law for magnetism,}\\
\nb\cross\E & =\bs 0, &  & \qquad\text{Faraday's law,}\\
\nb\cross\H & =\J+\frac{\bdry\D}{\bdry t}, &  & \qquad\textrm{{Ampere's\ law.}}
\end{alignat}
Constitutive relations between the various fields should be added
in order to make the system of equations solvable in principle.

We start by demonstrating in Section \ref{sec:Magnetostatics} how
magnetostatics, for which the Maxwell equations assume a particularly
simple form, may be obtained from the equation for the mechanical
power in continuum mechanics for the generalized case where the stress
tensor is antisymmetric, rather that symmetric.

Section \ref{sec:Manifolds-and-Differential} further motivates the
usefulness of antisymmetric tensors by presenting the full set of
Maxwell's equation using antisymmetric tensors\textemdash differential
forms\textemdash in a $4$-dimensional spacetime devoid of metric
properties. Then, some basic properties and operations corresponding
to differential forms are briefly reviewed.

Next, starting with the classical definition, it is shown in Section
\ref{sec:fluxes} how the notion of a flux vector field can be generalized
naturally and independently of a metric to the case of $n$-dimensional
manifolds by using differential $(n-1)$-forms.

In Section \ref{sec:Forces-and-stresses}, we consider the metric-invariant
geometric theory of smooth force and stress fields in generalized
media. In particular, a stress field is introduced as a tensor field
that acts on generalized velocity fields to produce $(n-1)$-forms
that model the flux of power.

Finally, as our main example, we show in Section \ref{sec:Maxwell's-Equations}
that the generalization of Maxwell's equations to $p$-form, metric-independent
electrodynamics in an $n$-dimensional spacetime results from the
basic properties of the stress under suitable assumptions. Specifically,
it is assumed th\,at a generalized velocity is a $p$-form, $\ga$,
in spacetime. Such a $p$-form is interpreted as a generalized vector
potential of electrodynamics. Next, it is assumed that the body force
vanishes. Then, a particularly simple form for the stress is assumed.
Namely, it is assumed that the stress is represented by an $(n-p-1)$-form
$\maxw$ so that the action of the stress $\st$ on the potential
form $\ga$ is given as
\begin{equation}
\st(\ga)=\maxw\wedge\ga,
\end{equation}
where a wedge denotes the exterior product of differential forms.

It is noted that antisymmetric tensors were used in the formulation
of electrodynamics by Truesdell and Toupin \cite[Section F]{Truesdell1960}.
Whittaker \cite[pp. 192--196]{Whittaker1953}, attributes the first
presentation of metric-free electrodynamics to Kottler \cite{Kottler1922},
while Truesdell and Toupin attribute the main contribution to van
Dantzig \cite{Dantzig1934}.

For a comprehensive introduction to geometric continuum mechanics,
which also includes a global non-smooth formulation, see \cite{Segev2023}.

\section{\label{sec:Magnetostatics}Magnetostatics and Generalized Media}

When we specialize the Maxwell equations to the case of magnetostatics,
we are concerned with the magnetic fields $\B$ and $\H$ in the case
where all fields are time-independent. Thus, the resulting equations
for magnetostatics are 
\begin{equation}
\nb\scp\B=0,\qquad\nb\cross\H=\J.\label{eq:magn-stat}
\end{equation}

Assuming that the physical space (at each instant) is simply connected,
the equation $\nb\scp\B=\bs 0,$implies that there is a (nonunique)
vector field $\A$, the \emph{vector potential, }such that 
\begin{equation}
\B=\nb\cross\A.\label{eq:v-pot}
\end{equation}
The identity $\nb\scp(\nb\cross\vct v)=0$, for every vector field
$\vct v$, implies that the divergence of the vector field $\B$,
as obtained from the vector potential, indeed vanishes.

Similarly, the equation $\nb\cross\H=\J$ implies that
\begin{equation}
\nb\scp\J=0\label{eq:cons-charge}
\end{equation}
\textemdash the conservation of charge. Conversely, conservation of
charge implies that there is a vector potential $\H$ such that $\J=\nb\cross\H$.
Thus, Equations (\ref{eq:v-pot},\ref{eq:cons-charge}) may be used
alternatively to (\ref{eq:magn-stat}).

As a motivation for the more general construction, we show in this
section that the equations of magnetostatics may be obtained as the
equations of the mechanics of a generalized continuum.

We recall that given the body force, $\bfc$, and the surface force,
$\sfc$, for the image in space of particular configuration of a material
body, $\reg$, the power expended for a virtual velocity $\bw$ ,
is given by
\begin{equation}
P_{\reg}(\vct{\vf})=\int_{\reg}\vct{\bfc}\scp\vct{\vf}\,\dV+\int_{\bdry\reg}\sfc\scp\vct{\vf}\,\dA.
\end{equation}

Consider the case whee $\bfc=\bs 0$, and use the basic property of
the stress tensor as expressed by the Cauchy formula, 
\begin{equation}
\sfc=\vct{\st}(\nor),
\end{equation}
where $\st$ is the Cauchy stress and $\nor$ is the unit normal to
$\bdry\reg$. The expression for the power assumes the form
\begin{equation}
P_{\reg}(\vct{\vf})=\int_{\bdry\reg}\vct{\st}(\nor)\scp\vct{\vf}\,\dA.
\end{equation}
Using the definition of the transpose of a linear mapping in a Euclidean
space, $\vct{\st}(\nor)\scp\vct{\vf}=\bs{\st}\trps(\bw)\scp\nor$
and the expression for the power becomes
\begin{equation}
P_{\reg}(\vct{\vf})=\int_{\bdry\reg}\bs{\st}\trps(\bw)\scp\nor\,\dA.
\end{equation}

For classical continuum mechanics, the Cauchy stress tensor is symmetric
so that the transposition is irrelevant. However, we may consider
the case where the stress is an antisymmetric tensor. Clearly, this
situation is outside the realm of classical continuum mechanics and
belongs to the mechanics of generalized media. Thus, it is assumed
henceforth that 
\begin{equation}
\bs{\st}\trps=-\bs{\st}.
\end{equation}

Since $\bs{\gs}$is assumed to be antisymmetric, it may be represented
by an axial vector $\H$, the components of which are given by
\begin{equation}
H_{p}=\shalf\eps_{pjk}\st_{jk}.
\end{equation}
In terms of $\H$, 
\begin{equation}
\bs{\st}\trps(\bw)=-\bs{\st}(\bw)=\H\cross\bw,
\end{equation}
and we can write for the power
\begin{equation}
P_{\reg}(\vct{\vf})=\int_{\bdry\reg}(\H\cross\bw)\scp\nor\,\dA.
\end{equation}
Using the Gauss theorem,
\begin{equation}
P_{\reg}(\vct{\vf})=\int_{\reg}\nb\scp(\H\cross\bw)\dV.
\end{equation}
We now make use of the identity
\begin{equation}
\nb\scp(\H\cross\bw)=(\nb\cross\H)\scp\bw-\H\scp(\nb\cross\bw),\label{eq:eq:identity}
\end{equation}
and obtaint
\begin{equation}
P_{\reg}(\vct{\vf})=\int_{\reg}(\nb\cross\H)\scp\bw\dV-\int_{\reg}\H\scp(\nb\cross\bw)\dV.\label{eq:power-}
\end{equation}

Anticipating the analogy with magnetostatics, we change the notation
$\bw$ to $\A$ to arrive at
\begin{equation}
P_{\reg}(\A)=\int_{\reg}(\nb\cross\H)\scp\A\dV-\int_{\reg}\H\scp(\nb\cross\A)\dV.
\end{equation}
Moreover, we may define 
\begin{equation}
\J:=\nb\cross\H,\qquad\B:=\nb\cross\A,
\end{equation}
which implies immediately that
\begin{equation}
\nb\scp\J=0,\qquad\nb\scp\B=0.
\end{equation}
In terms of $\J$ and $\B$, the expression for the power is
\begin{equation}
P_{\reg}(\A)=\int_{\reg}\J\scp\A\dV-\int_{\reg}\H\scp\B\dV.
\end{equation}

Evidently, we may now reinterpret $\A$ as the vector potential, $\J$
as the current density, $\H$ as the magnetic field intensity, and
$\B$ as the magnetic flux density. In other words, if we admit antisymmetric
stress tensors, which in the mechanical interpretation may be viewed
as a smooth distribution of moments, mechanics has the same mathematical
structure as magnetostatics.

It is noted that, for continuum mechanics, the power may also be written
as
\begin{equation}
P_{\reg}(\vct{\vf})=\int_{\reg}\st_{ij}\vf_{i,j}\dV.
\end{equation}
where the summation convention is used, and a comma denotes partial
differentiation. Thus, for the case of an antisymmetric stress tensor,
only the antisymmetric, spin part $(\vf_{i,j}-\vf_{j,i})/2$, of the
velocity gradient is relevant. This was expected as we consider a
generalized continuum for which the kinematics is not represented
merely by the velocity of the material points in space. 

\section{\label{sec:Manifolds-and-Differential}Manifolds and Differential
Forms}

It is observed that the field equations for magnetostatics, as presented
above, look simpler and more symmetric than the full set of Maxwell's
equations. However, in an appropriate $4$-dimensional setting, the
Maxwell equations attain the same simplicity and symmetry as the equations
for magnetostatics. In fact, they are completely analogous.

In addition to the simplicity, the version of the Maxwell equations
we write below, does not use any metric properties of the ambient
medium, be it the physical space or a body of continuum mechanics.
Thus, as geometrical objects, the electromagnetic fields in a body
are independent of deformations of the body in space. In addition,
this may be relevant for general relativity, where the metric tensor
is unknown a priori.

The Faraday $2$-tensor $\frdy$ and the Maxwell $2$-tensor $\maxw$
are defined, respectively, by
\begin{equation}
\{\frdy_{ij}\}=\begin{pmatrix}\begin{array}{llll}
0 & -E_{1} & -E_{2} & -E_{3}\\
E_{1} & \hphantom{-}0 & \hphantom{-}B_{3} & -B_{2}\\
E_{2} & -B_{3} & \hphantom{-}0 & \hphantom{-}B_{1}\\
E_{3} & \hphantom{-}B_{2} & -B_{1} & \hphantom{-}0
\end{array}\end{pmatrix},\quad\text{and}\quad\{\maxw_{ij}\}=\begin{pmatrix}\begin{array}{llll}
0 & -H_{1} & -H_{2} & -H_{3}\\
H_{1} & \hphantom{-}0 & \hphantom{-}D_{3} & -D_{2}\\
H_{2} & -D_{3} & \hphantom{-}0 & \hphantom{-}D_{1}\\
H_{3} & \hphantom{-}D_{2} & -D_{1} & \hphantom{-}0
\end{array}\end{pmatrix}.
\end{equation}
In addition, consider the vector
\begin{equation}
\{\wh{\fflow}\vphantom{\fflow}^{i}\}=\begin{pmatrix}\begin{array}{l}
\rho\\
J^{1}\\
J^{2}\\
J^{3}
\end{array}\end{pmatrix},
\end{equation}
\emph{i.e., }$\wh{\fflow}\vphantom{\fflow}^{1}=\rho$, $\wh{\fflow}\vphantom{\fflow}^{2}=J^{1}$,
etc., and define the antisymmetric $3$-tensor $\fflow_{jkl}$ by
\begin{equation}
\fflow_{jkl}:=\wh{\fflow}\vphantom{\fflow}^{i}\eps_{ijkl}.
\end{equation}

Similarly to the equations of magnetostatics, the Maxwell equations
may be written now in the form
\begin{equation}
\dee\frdy=0,\qquad\dee\maxw=\fflow,\label{eq:maxw-4-d}
\end{equation}
where $\dee$ denotes a differential operator to be described below.
If the potential $1$-tensor $\ga$ is defined by
\begin{equation}
\{\ga_{i}\}=(\begin{array}{cccc}
\phi & A_{1} & A_{2} & A_{3}\end{array}),
\end{equation}
where $\phi$ is the electrostatic potential, then, the analogs of
Equations (\ref{eq:v-pot},\ref{eq:cons-charge}) are, respectively,
\begin{equation}
\frdy=\dee\ga,\quad\text{and}\quad\dee\fflow=0.\label{eq:Alternatives}
\end{equation}

Now that we have demonstrated the power provided by making use of
antisymmetric tensors, we will shortly and roughly review the main
definitions and results. For details, see \cite{Segev2023} or the
comprehensive \cite{Abraham1988,Lee2002}.

Differentiable manifolds are characterized by the property that points
on them can be identified using local coordinate systems, or charts,
with points in $\rn$. Transformations of coordinates are smooth and
reversible to make the assignment of coordinates consistent. Smooth
curves are well-defined on a manifold. A \emph{tangent vector} at
a point may be defined as the derivative of a smooth curve passing
through that point with respect to the curve parameter, and one writes
\begin{equation}
v=\left.\frac{\dee c}{\dee t}\right|_{t=0},
\end{equation}
where $c(t)$ is the curve.

An example of a differentiable manifold is the configuration space
modeling the kinematics of a mechanical system having a finite number
of degrees of freedom. Another example is a body of continuum mechanics
for which a natural reference configuration is not given. For this
example, viewing a tangent vector as an infinitesimal vector adjoining
two neighboring points, it is evident that two tangent vectors at
distinct points cannot be compared, added, etc. as objects should
be invariant under any superimposed deformation. If $\base$ denotes
a differentiable manifold, the collection of tangent vectors at an
arbitrary point $x\in\base$, denoted by $T_{x}\base$, may be given
the structure of a vector space. In general, no inner product structure
is given on $T_{x}\base$. 

At each point, one can consider completely antisymmetric $r$-tensors,
scalar-valued $r$-multilinear mappings of tangent vectors at this
point. Such a tensor field, $\go$ assigning an antisymmetric tensor
$\go(x)$ on $T_{x}\base$ for each $x\in\base$, is a \emph{differential
$r$-form.} Thus, the electromagnetic fields introduced above are
differential forms. For example, the field $\fflow$ is a differential
$3$-form on the $4$-dimensional spacetime.

Any tensor $T$, in general not antisymmetric, induces an antisymmetric
tensor $\alt(T)$. The components of $\alt(T)$ may be obtained from
the components of $T$ using the alternating symbol. The operation
$\alt$ is a projection in the sense that an antisymmetric tensor
is invariant under its action.
\begin{rem}
\label{rem:dimension}Note that the components $\go_{i_{1}\dots i_{r}}$,
$i_{m}=\oneto n$, $m=\oneto r$, of any completely antisymmetric
$r$-tensor in an $n$-dimensional space are determined by the components
$\go_{\ggm_{1}\dots\ggm_{r}}$, $\ggm_{m}=\oneto n$, $m=\oneto r$,
with the properties that $\ggm_{m+1}>\ggm_{m}$. It follows that the
dimension of the space of antisymmetric $r$-tensors in an $n$-dimensional
space is 
\begin{equation}
\dimension(r)=\frac{n!}{r!(n-r)!},\label{eq:dim-anti}
\end{equation}
the number of possible ways to choose $r$ numbers from $\oneto n$.
In particular,
\begin{equation}
\dimension(0)=\dimension(n)=1,\qquad\dimension(1)=\dimension(n-1)=n.
\end{equation}
\end{rem}

The \emph{exterior product }of an $r$-form $\go$ and a $p$-form
$\gf$, is the $(r+p)$ form $\go\wedge\gf$ defined by
\begin{equation}
\go\wedge\gf=\frac{(r+p)!}{r!p!}\alt(\go\otimes\gf).
\end{equation}
Thus, in essence, the exterior product is the antisymmetrized tensor
product. In fact, the exterior product generalizes the ``cross'' product
of vectors in a $3$-dimensional Euclidean space to any pair of tensors
in any dimension of the ambient space without using an inner product.

The \emph{exterior derivative}, $\dee\go$, of a differential $r$-form,
$\go$, can roughly by described as the antisymmetrized gradient of
the differential form. It is an $(r+1)$-form that has the following
properties. 

For a $0$-form, a real valued function $f$ defined on the manifold,
the $1$-form $\dee f$ is defined by
\begin{equation}
\dee f(v):=\frac{\dee}{\dee t}(f(c(t)))\resto{t=0},
\end{equation}
where $v=\dee c/\dee t\resto{t=0}$. It follows that $\dee f(v)$
is the directional derivative of the function in the direction of
$v$.

The second property of exterior differentiation is that
\begin{equation}
\dee^{2}\go=\dee(\dee\go)=0.\label{eq:d^2=00003D0}
\end{equation}
Roughly, as the second derivative is symmetric its antisymmetrization
naturally vanishes. This property generalizes classical identities
of vector analysis, such as $\nb\scp(\nb\cross\vct v)=0$ and $\nb\cross(\nb f)=\bs 0$.

Finally, for an $r$-form $\go$ and a $p$-form $\gf$,
\begin{equation}
\dee(\go\wedge\gf)=\dee\go\wedge\gf+(-1)^{r}\go\wedge\dee\gf,\label{eq:d(a^b)}
\end{equation}
which generalizes the Leibniz rule.

These three properties are sufficient to define the exterior derivative
uniquely as a linear differential operator. In terms of components,
the exterior derivative $\dee\go$ may be expressed as the exterior
product of the operator $\partial/\partial x^{i}$ with the representative
of $\go$.

As mentioned above, for a function, a $0$-form, the exterior derivative
is analogous to the gradient. The exterior derivative of a $1$-form
is analogous to the curl of a vector field. The exterior derivative
of a $2$-form in a $3$-dimensional manifold, or in general, the
exterior derivative of an $(n-1)$-form in an $n$-dimensional manifold,
is analogous to the divergence of a vector field.

\medskip{}

\begin{figure}
\begin{centering}
\includegraphics{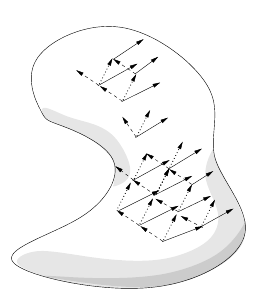}
\par\end{centering}
\caption{\label{fig:integration}Illustrating integration of a differential
form}

\end{figure}

Differential forms are natural integrands on manifolds. The integral
of an $n$-form over an $n$-dimensional manifold may be described
roughly as follows. Divide the manifold into infinitesimal parallelepipeds,
each of which is determined by a collection of $n$ tangent vectors.
Here, because of the antisymmetry of the differential form to be integrated,
one has to make sure that the collections of vectors at the various
points are ordered consistently. Then, add up the actions of the
differential form on the collections of vectors determining the parallelepipeds
at the various points (see a rough illustration in Figure \ref{fig:integration}).

The integration of the $n$-form, $\go$, over the $n$-dimensional
manifold, $\base$, is denoted as
\begin{equation}
\int_{\base}\go.
\end{equation}

Using the language of differential forms and their integration, the
classical Green theorem, Gauss theorem, and Kelvin-Stokes theorem,
may be generalized to one compact form. Specifically, the geometric
Stokes theorem states that for an $(n-1)$-form $\go$, the integral
over the boundary $\bdry\base$ of an $n$-dimensional manifold $\base$,
satisfies
\begin{equation}
\int_{\bdry\base}\go=\int_{\base}\dee\go.\label{eq:stokes}
\end{equation}
Note that $\bdry\base$ is an $(n-1)$-dimensional manifold and $\dee\go$
is an $n$-form, so that the integrals above are meaningful. The theorem
is based on the fundamental theorem of calculus.

\section{\label{sec:fluxes}The Representation of Fluxes by Differential Forms}

The notion of a flux vector field, such as the heat flux vector field,
or the momentum flux vector field, is fundamental in continuum mechanics.
We will denote a generic flux vector field of classical continuum
mechanics by $\fv$. If $\subman$ is a surface in a $3$-dimensional
Euclidean with unit normal vector field $\nor$, the flux through
$\subman$, in the sense determined by $\nor$, is given by 
\begin{equation}
\Phi=\int_{\subman}\fv\scp\nor\dA.\label{eq:flux-classic}
\end{equation}
The infinitesimal version is evidently
\begin{equation}
\dee\Phi=\fv\scp\nor\dA=\fv\scp\dee\vct A,
\end{equation}
where $\dee\vct A:=\dee A\nor$ is the infinitesimal area vector so
that $\dee A=\abs{\dee\vct A}$.

Consider the case where the infinitesimal area vector is determined
by two infinitesimal vectors $\vct v_{1}$ and $\vct v_{2}$ originating
from a point $y$ on the surface so that
\begin{equation}
\dee\vct A=\dee A\nor=\vct v_{1}\cross\vct v_{2}.
\end{equation}
Then, the infinitesimal flux through the area element is given by
\begin{equation}
\dee\Phi=\fv\scp(\vct v_{1}\cross\vct v_{2}).
\end{equation}

We now view $\dee\Phi$ as the evaluation of a function $\go_{y}$
of the two vectors and write this as
\begin{equation}
\dee\Phi=\go_{y}(\vct v_{1},\vct v_{2})=\fv\scp(\vct v_{1}\cross\vct v_{2}).
\end{equation}
By its definition and the properties of the ``cross'' product, $\go_{y}$
is linear in each of the vectors and is antisymmetric. It is concluded
that $\go_{y}$ is an antisymmetric $2$-tensor and assuming it varies
smoothly with $y$, there is a $2$-form $\go$ such that
\begin{equation}
\dee\Phi=\go(y)(\vct v_{1},\vct v_{2})\qquad\text{and}\qquad\Phi=\int_{\subman}\go.\label{eq:flux-forms}
\end{equation}

It is emphasized that while the classical expression for the flux
as in Equation (\ref{eq:flux-classic}) depends on the availability
of the scalar product, the expressions in Equation (\ref{eq:flux-forms})
in terms of differential forms are independent of any scalar product
and remain invariant under deformations of $\subman$. 

In general, one can use the integration of an $r$-form on an $r$-dimensional
submanifold as a generalization of the standard flux in three dimensions.
In most cases, $r=n-1$. For example, the spacetime of classical mechanics
is a $4$-dimensional manifold without a natural scalar product. Integration
of forms enables one to compute fluxes of $3$-forms through $3$-dimensional
submanifolds of spacetime and also the fluxes of $2$-forms through
$2$-dimensional submanifolds. 

\section{\label{sec:Forces-and-stresses}Forces and Traction Stresses on Manifolds}

We have already emphasized that tangent vectors at distinct points
in a manifold cannot be compared, added, etc. As a consequence, the
same applies to tensors acting on vectors. It follows that it is meaningless
to integrate force vectors over a body in the setting of a manifold.
As noted above, natural integrands on an $n$-dimensional manifolds
are $n$-forms, which we interpret as densities of some scalar extensive
properties.

For the case of forces, the density to be integrated is that of the
mechanical power expended by a force field for a given velocity field.
Consequently, the notion of mechanical power assumes a major role
in the formulation of metric-invariant force and stress theory.

Consider, for example, body forces in continuum mechanics. For an
infinitesimal volume element $\dV$, the power expended by the body
force $\bfc(x)$ for the velocity $\bw(x)$ is 
\begin{equation}
\dee P=\bfc(x)\cdot\bw(x)\dV.
\end{equation}
Let $\vct v_{1},\,\vct v_{2},\,\vct v_{3}$ be three linearly independent
infinitesimal vectors that generate a parallelepiped with volume $\dV$,
so,
\begin{equation}
\dV=\vct v_{1}\scp(\vct v_{2}\cross\vct v_{3})=\det[\vct v_{1},\,\vct v_{2},\,\vct v_{3}].
\end{equation}
Thus,
\begin{equation}
\dee P=\bfc(x)\cdot\bw(x)\det[\vct v_{1},\,\vct v_{2},\,\vct v_{3}].
\end{equation}
We wish to view $\dee P$ as a function $\go_{\bw}(x)$ of the vectors
$\vct v_{1},\,\vct v_{2},\,\vct v_{3}$ so
\begin{equation}
\dee P=\go_{\bw}(x)(\vct v_{1},\vct v_{2},\vct v_{3})=\bfc(x)\cdot\bw(x)\det[\vct v_{1},\,\vct v_{2},\,\vct v_{3}].
\end{equation}
The properties of the determinant imply immediately that $\go_{\bw}(x)$
is an antisymmetric $3$-tensor and so, assuming that it varies smoothly
with $x$, $\go_{\bw}$ is a $3$-form. In addition, $\go_{\bw}(x)$
depends linearly on $\bw(x)$. 

Hence, the body force at the point $x$ is a linear mapping acting
on values of velocity fields to produce antisymmetric $3$-tensors.
The body force field is a tensor field $\bfc$ that when acting on
a velocity field, it yields a $3$-form $\bfc(\bw)$. It is observed
that the definition extends naturally to the case of a generalized
medium, where $\bw$ is any generalized velocity field. This definition
does not require any metric property of space and the body force field,
so defined, is invariant under deformations of the body. This definition
also generalizes naturally to any dimension by replacing $3$ above
with a generic $n$. The total power of the body force acting on a
region $\reg$ is given, therefore, as an integral of the resulting
$n$-form by
\begin{equation}
P_{\bfc}=\int_{\reg}\bfc(\bw).
\end{equation}

\medskip{}

Analogous arguments imply that the surface force, $\sfc_{\reg}$ acting
on the boundary $\bdry\reg$, is a tensor field that acts on generalized
velocity fields, $\bw$, to produce an $(n-1)$-forms, $\sfc_{\reg}(\bw)$,
defined on $\bdry\reg$, and the power expended is
\begin{equation}
P_{\sfc}=\int_{\bdry\reg}\sfc_{\reg}(\bw).
\end{equation}
It is emphasized that for $y\in\bdrr$, $\sfc_{\reg}(y)(\bw(y))$
is an antisymmetric tensor acting on $n-1$ vectors that are tangent
to $\bdrr$ at $y$.

\medskip{}

The considerations corresponding to the stress object combine those
related to fluxes with those related to forces. For standard continuum
mechanics, the Cauchy stress $\st(x)$, at a point $x$ in the body,
determines the surface force on any subbody $\reg$, the boundary
of which contains $x$, using the Cauchy formula
\begin{equation}
\sfc_{\reg}(x)=\st(x)(\nor(x)),
\end{equation}
where $\nor(x)$ is the unit normal to $\bdrr$ at $x$.

Let $\dA$ be an infinitesimal area element on $\bdrr$ at $x$. Then
the infinitesimal power flux through $\dA$ for a velocity $\bw(x)$
is
\begin{equation}
\dee P_{\sfc}=\sfc_{\reg}(x)\scp\bw(x)\dA=\st(x)(\nor(x))\scp\bw(x)\dA.
\end{equation}
Using the definition of the transpose of a linear mapping, 
\begin{equation}
\st(x)(\nor(x))\scp\bw(x)=\st(x)\trps(\bw(x))\scp\nor(x),
\end{equation}
so,
\begin{equation}
\dee P_{\sfc}=\st(x)\trps(\bw(x))\scp\nor(x)\dA.
\end{equation}
For standard continuum mechanics, the stress tensor is symmetric so
that transposition has no consequences. However, anticipating the
extension to generalized media, we keep the indication of a transposition.

Let $\vct v_{1},\,\vct v_{2}$ be two infinitesimal vectors tangent
to $\bdrr$ at $x$ that induce a parallelogram tangent to $\bdrr$
of area 
\begin{equation}
\dA=\abs{\vct v_{1}\cross\vct v_{2}},\quad\text{and}\quad\nor(x)\dA=\vct v_{1}\cross\vct v_{2}.
\end{equation}
Then, the power is given as
\begin{equation}
\dee P_{\sfc}=\st(x)\trps(\bw(x))\scp(\vct v_{1}\cross\vct v_{2}).
\end{equation}
As before, we wish to view the power as the evaluation of a function
$\psi_{\bw}$ on the vectors $\vct v_{1},\,\vct v_{2}$. That is,
\begin{equation}
\dee P_{\sfc}=\psi_{\bw}(x)(\vct v_{1},\vct v_{2})=\st(x)\trps(\bw(x))\scp(\vct v_{1}\cross\vct v_{2}).
\end{equation}
It is emphasized that in this expression the vectors $\vct v_{1},\,\vct v_{2}$
are viewed as vectors in the body and need not be tangent to the boundary
of a particular body, in contrast with the case of a surface force
on a particular subbody. Once the two vectors are given, the power
of the surface force $\sfc_{\reg}$, for any subbody the tangent plane
to the boundary of which contains these vectors, is induced. 

Comparing the last equation with the discussion in Section $\ref{sec:fluxes}$,
it follows that 
\begin{equation}
\psi_{\bw}=\st\trps(\bw),\qquad\psi_{\bw}(x)=\st\trps(x)(\bw(x)),
\end{equation}
is a $2$-form in the body that depends linearly on the values of
$\bw$.

It is concluded that the stress object is a tensor field that, when
acting on a velocity field it produces a power flux $2$-form in the
body. It is noted again that the antisymmetric $2$-tensor $\st(x)\trps(\bw(x))$
acts on any pair of vectors at $x$, and is not restricted, as $\sfc_{\reg}$
is, to vectors tangent to the boundary of a particular subbody. In
fact, the analog of the Cauchy formula is simply the restriction of
the tensor $\st(x)\trps(\bw(x))$ to vectors tangent to $\bdrr$.

The extension to generalized media in $n$ dimensions is straightforward.
The stress is a tensor field that acts on generalized velocity fields
to produce $(n-1)$-forms. Given such a stress $\st$, the power of
the surface force $\sfc_{\reg}$ induced by the stress on a subbody
$\reg$ is given by
\begin{equation}
P_{\sfc}=\int_{\bdrr}\st(\bw).
\end{equation}

The Stokes theorem implies immediately the power may also be written
as
\begin{equation}
P_{\sfc}=\int_{\reg}\dee(\st(\bw)).
\end{equation}

\section{\label{sec:Maxwell's-Equations}Maxwell's Equations and $p$-Form
Electrodynamics}

Finally, we show how metric-independent, $p$-form electrodynamics
in an $n$-dimensional spacetime results as a special case of geometric
stress theory as described above. To that end, we make the following
assumptions.

\subsubsection*{Assumption I}

Generalized velocity fields are $p$-forms in $n$-dimensional spacetime.

A $p$-form representing a generalized velocity is interpreted as
a potential for the electromagnetic fields. This interpretation will
be justified below. In order to reflect this interpretation, we will
change the notation for a generic generalized velocity form $\bw$
to $\ga$.

It is recalled that the value of the stress at a point is a linear
mapping that acts on a generalized velocity\textemdash now, an antisymmetric
$p$-tensor\textemdash to yield an antisymmetric $(n-1)$-tensor representing
the flux of power. In light of Remark \ref{rem:dimension}, the dimension
of the space of antisymmetric $p$-tensors is 
\begin{equation}
\dimension(p)=\frac{n!}{p!(n-p)!},
\end{equation}
and that corresponding to antisymmetric $(n-1)$-tensors is $n$.
Hence, the dimension of the space of values of the stress tensor at
a point is
\begin{equation}
\dimension(p)\dimension(n-1)=\frac{n!}{p!(n-p)!}n.\label{eq:dim-stress-gen}
\end{equation}

\subsubsection*{Assumption II}

The body force vanishes, that is, $\bfc=0$.

As a result of this assumption, the total power for a given generalized
velocity $\ga$ is
\begin{equation}
P=\int_{\bdrr}\st(\ga)=\int_{\reg}\dee(\st(\ga)).
\end{equation}
The most significant assumption is

\subsubsection*{Assumption III}

There is an $(n-p-1)$-form, $\maxw$, such that the power flux $(n-1)$-form,
$\st(\ga)$, is given by
\begin{equation}
\st(\ga)=\maxw\wedge\ga.
\end{equation}

It is noted first that the degrees of the forms $\maxw$ and $\ga$
indeed add up to $n-1$. It is also noted that this assumption imposes
a significant restriction on the stress. The dimension of the space
of values of antisymmetric $(n-p-1)$-tensors is
\begin{equation}
\dimension(n-p-1)=\frac{n!}{(n-p-1)!(n-(n-p-1))!}=\frac{n!}{(n-p-1)!(p+1)!},
\end{equation}
in comparison with the dimension for the general case given by Equation
(\ref{eq:dim-stress-gen}).

It follows from this assumption that the power is given by
\begin{equation}
P=\int_{\bdry\reg}\maxw\wedge\ga=\int_{\reg}\dee(\maxw\wedge\ga).\label{eq:power-10}
\end{equation}
We may now use the rule for the differentiation of the exterior product,
(\ref{eq:d(a^b)}), to obtain
\begin{equation}
P=\int_{\reg}\dee\maxw\wedge\ga+(-1)^{n-p-1}\int_{\reg}\maxw\wedge\dee\ga.
\end{equation}
Setting,
\begin{equation}
\frdy=\dee\ga\quad\text{and}\quad\fflow=\dee\maxw,\label{eq:def-fields}
\end{equation}
the expression for the power assumes the form
\begin{equation}
P=\int_{\reg}\fflow\wedge\ga+(-1)^{n-p-1}\int_{\reg}\maxw\wedge\frdy.
\end{equation}
Finally, it follows immediately from the definitions in (\ref{eq:def-fields})
and property (\ref{eq:d^2=00003D0}) of differential forms that 
\begin{equation}
\dee\frdy=0,\quad\text{and}\quad\dee\fflow=0.
\end{equation}

\bigskip{}

\noindent \textbf{\textit{Acknowledgments.}} This work was partially
supported by the Pearlstone Center for Aeronautical Engineering Studies
at Ben-Gurion University.


\end{document}